\documentclass[conference]{IEEEtran}
\IEEEoverridecommandlockouts
\usepackage{cite}
\usepackage{amsmath,amssymb,amsfonts}
\usepackage{algorithmic}
\usepackage{graphicx}
\usepackage{textcomp}
\usepackage{xcolor}
\usepackage{hyperref}
\usepackage{balance}
\usepackage{booktabs}

\def\BibTeX{{\rm B\kern-.05em{\sc i\kern-.025em b}\kern-.08em
    T\kern-.1667em\lower.7ex\hbox{E}\kern-.125emX}}

\begin{document}
\title{A Lightweight Neural Network for Accelerating Radiative Transfer Modeling in WRF \\
{\footnotesize \textsuperscript{*}}
\thanks{}
}

\author{
\IEEEauthorblockN{
    Erick Fredj\IEEEauthorrefmark{1}\IEEEauthorrefmark{2}\thanks{Corresponding author: fredj@g.jct.ac.il}, 
    Iggy Segev Gal\IEEEauthorrefmark{2}, 
    Noam Lavi\IEEEauthorrefmark{2}, \\
    Shahar Belkar\IEEEauthorrefmark{2}, 
    Mark Wasserman\IEEEauthorrefmark{2}, 
    Ding Zhaohui\IEEEauthorrefmark{2}, 
    Yann Delorme\IEEEauthorrefmark{2}
}
\vspace{1\baselineskip}
\IEEEauthorblockA{\IEEEauthorrefmark{1}Department of Computer Science, The Jerusalem College of Technology, Jerusalem, Israel.\\ Email: fredj@g.jct.ac.il}
\IEEEauthorblockA{\IEEEauthorrefmark{2}Toga Networks, a Huawei Company, Tel Aviv, Israel}
}



      




\maketitle

\begin{abstract}
Radiative transfer calculations in weather and climate models are notoriously complex and computationally intensive, which poses significant challenges. Traditional methods, while accurate, can be prohibitively slow, necessitating the development of more efficient alternatives. Recently, empirical emulators based on neural networks (NN) have been proposed as a solution to this problem. These emulators aim to replicate the radiation parametrization used in the models, at a fraction of the computational cost.
However, a common issue with these emulators is that their accuracy has often been insufficiently evaluated, especially for extreme events for which the amount of training data is sparse. The current study proposes such a model for accelerating radiative heat transfer modeling in WRF, and validates the accuracy of the approach for an extreme weather scenario.
\end{abstract}

\begin{IEEEkeywords}
Surrogate Models, Weather Research and Forecasting (WRF), Neural Network, Optimization, Machine Learning
\end{IEEEkeywords}

\section{Introduction}
\label{section:introductions}
The computation of the Rapid Radiative Transfer Model for General Circulation Models (RRTMG) within the Weather Research and Forecasting (WRF) model constitutes a substantial portion of the total computational effort, ranging from up to 20\% on clear sky days to as much as 60\% on cloudy days. Although the specific percentage varies depending on the model configuration, radiative transfer calculations are particularly time-consuming components of WRF models (e.g., \cite{Morcrette2007}, \cite{Morcrette2008}; \cite{Manners2009}). In both climate modeling and numerical weather prediction (NWP), there is an inherent trade-off between the accuracy and computational efficiency of radiative transfer calculations. Highly accurate methods, such as line-by-line procedures, could theoretically compute radiative fluxes at every grid point and every time step. However, performing radiative transfer calculations at this level of detail would require as much or more CPU time than all other model components combined, including model dynamics and other physical parametrization \cite{Morcrette2008}. To manage this computational load, various simplifications are often implemented.

One common approach is the correlated-k method (Lacis and Oinas, 1991), which reduces the integration over wavelength by grouping wavelengths with similar absorption coefficients (k terms). This significantly decreases the number of monochromatic radiative transfer calculations needed. The number of k terms can be adjusted to balance accuracy and efficiency for a specific application; however, the correlated-k methods are not efficient enough to allow calculations at every grid point and time step.

To further reduce computational costs, radiative calculations are typically performed at reduced temporal and/or spatial resolutions. For example, in climate and global forecast models at the National Centers for Environmental Prediction (NCEP) and the United Kingdom Met Office (UKMO), radiative calculations are performed every 1 to 3 hours \cite{Manners2009}. This means that significant changes in radiative profiles, due to factors like cloud cover and the angle of incident solar radiation, might not be captured between these calculations. At the European Center for Medium-Range Weather Forecasts (ECMWF), radiative calculations are conducted on a coarser grid and then interpolated to a finer grid, speeding up the calculations \cite{Morcrette2007}, \cite{Morcrette2008}. Similarly, the Canadian operational Global Environmental Multi-scale model uses a reduced vertical resolution approach, calculating full radiation at alternate vertical levels and interpolating for intermediate levels (e.g., \cite{Cote1998a},\cite{Cote1998b}). These strategies reduce the horizontal or vertical variability in radiation fields and can compromise the accuracy and consistency of the model's radiation calculations, potentially affecting climate simulations and weather predictions.

The trade-off between computational speed and accuracy in radiation calculations has led to the development of alternative numerical algorithms designed to enhance computational efficiency while maintaining accuracy. Two techniques proposed to improve the temporal and spatial resolution of radiation calculations involve enhanced interpolation methods for radiative calculations from a coarse to a fine grid \cite{Morcrette2008} and improved radiative calculations between full-time steps ( \cite{Venema2007}; \cite{Manners2009} ). However, these approaches can introduce significant errors over time due to accumulated discrepancies in interactions with other dynamic and physical processes (\cite{Emanuel2004}).

Given the need to balance speed and accuracy, data-driven radiation emulators based on neural networks (NNs) have emerged as promising alternatives ( \cite{Song2021}; \cite{Song2022}; \cite{Song2020} ), offering significant improvements in computational speed while maintaining reasonable accuracy. Chevallier et al. ( \cite{Chevallier1998}, \cite{Chevallier2000} ) pioneered NN-based longwave (LW) radiation emulation for ECMWF models. Since then, NN-based LW and shortwave (SW) emulators have been developed for various models, including the Community Atmosphere Model (CAM), the Climate Forecast System (CFS), and the Super-Parameterized Energy Exascale Earth System Model (SP-E3SM) ( \cite{Belochitski2021}; \cite{Krasnopolsky2005}; \cite{Krasnopolsky2008}; \cite{Krasnopolsky2010}; \cite{Pal2019} ). Krasnopolsky et al. \cite{Krasnopolsky2010} reported impressive results for an NN-based emulator of the Rapid Radiative Transfer Model for General Circulation Models (RRTMG; \cite{Clough2005}; \cite{Iacono2008} ), achieving computational speed improvements of 16-60 times while preserving long-term (17-year) stability. In Belochitski et al. \cite{Belochitski2021}, the NN-based radiation emulator outperformed those based on Classification and Regression Trees (CART). Recently, Pal et al. \cite{Pal2019} achieved a tenfold speed increase and 90-95 \% accuracy using a deep neural network (DNN), despite the greater computational demands of DNNs. Various emulators have also been developed for idealized frameworks (e.g.,  \cite{Krasnopolsky2014}; \cite{Rasp2018} ) and specific processes such as convection \cite{Gentine2018}, the planetary boundary layer \cite{Wang2019}, and dynamics.  While advanced deep-learning techniques have been applied to post-processing in weather and climate models, emulating dynamic and physical processes within numerical models remains a challenging task.

In this study, we develop a new lightweight neural-network-based model for radiative heat transfer, and investigate its performance within the Weather Research and Forecasting (WRF) model.
We follow the approach of Krasnopolsky et al. \cite{Krasnopolsky2014} to develop a neural network (NN) emulator for both longwave and shortwave radiation physics formulations. The accuracy of the model is demonstrated for the challenging forecast of Typhoon Muifa, which made landfall near China’s largest metropolitan area and several major shipping ports on September 14, 2022.

The following sections provide a detailed description of the methodology. Section \ref{section:methods} details specifics about the modeling approach and data used to train the neural network. A qualitative and quantitative evaluation of the accuracy of the model, along with performance analysis, is given in Section \ref{section:resultsanddiscussions}. Finally, a summary and prospect for future research are presented in Section \ref{section:conclusions}.

\section{Modeling Approach}
\label{section:methods}
\subsection{Neural Network Emulator for RRTMG Model}
\label{subsection:NNModel}
This study proposes a light-weight neural network emulator for the Rapid Radiative Transfer Model for GCMs (RRTMG), composed of a single hidden layer. The proposed neural network includes 178 input variables for the clear-sky model, and 222 input variables for the cloudy-sky model. The output variables consist of 47 longwave (LW) and shortwave (SW) components, as shown in Table \ref{tab:input-output}.
\begin{table}[ht]
\centering
\begin{tabular}{|c|c|}
    \hline
    \multicolumn{2}{|c|}{Inputs (178,222)} \\ \hline
    Vertical Pressure & 44 \\ \hline
    Vertical Temperature & 44 \\ \hline
    Vertical Ozone & 44 \\ \hline
    Vertical Cloud fraction (cloudy) & 44 \\ \hline
    Skin temperature (LW) & 1 \\ \hline
    Surface emissivity (LW) & 1 \\ \hline
    Cosine solar zenith angle multiply by solar constant (SW) & 1 \\ \hline
    Surface albedo (SW) & 1 \\ \hline
    \multicolumn{2}{|c|}{Outputs (47)} \\ \hline
     Vertical total sky heating rate (LW,SW) & 44 \\ \hline
     Total sky long-wave upward flux at the top (LWUPT) & 1 \\ \hline
     Total sky long wave upward flux at the bottom (LWUPB) & 1 \\ \hline
     Total sky long-wave downward flux at the bottom (LWDNB)  & 1 \\ \hline
     Total sky shortwave upward flux at the top (SWUPT) & 1 \\ \hline
     Total sky shortwave downward flux at the bottom (SWDNB) & 1 \\ \hline
     Total sky shortwave upward flux at the bottom (SWUPT) & 1 \\ \hline
\end{tabular}
\caption{ List of Input and Output Variables for Neural Network Emulators of Longwave (LW) and Shortwave (SW) Radiation }
\label{tab:input-output}
\end{table}
The proposed model follows an 8-category approach proposed by Song and Roh \cite{Song2021}, whereby 8 sub-models are trained independently for combinations of the following conditions: Long-wave or short-wave radiation, clear or cloudy skies, and over land or ocean. Each sub-model is trained on its respective data. This method aimed to improve data utilization and reduce representation errors.

\begin{figure}
	\centerline{\includegraphics[width=3.5in]{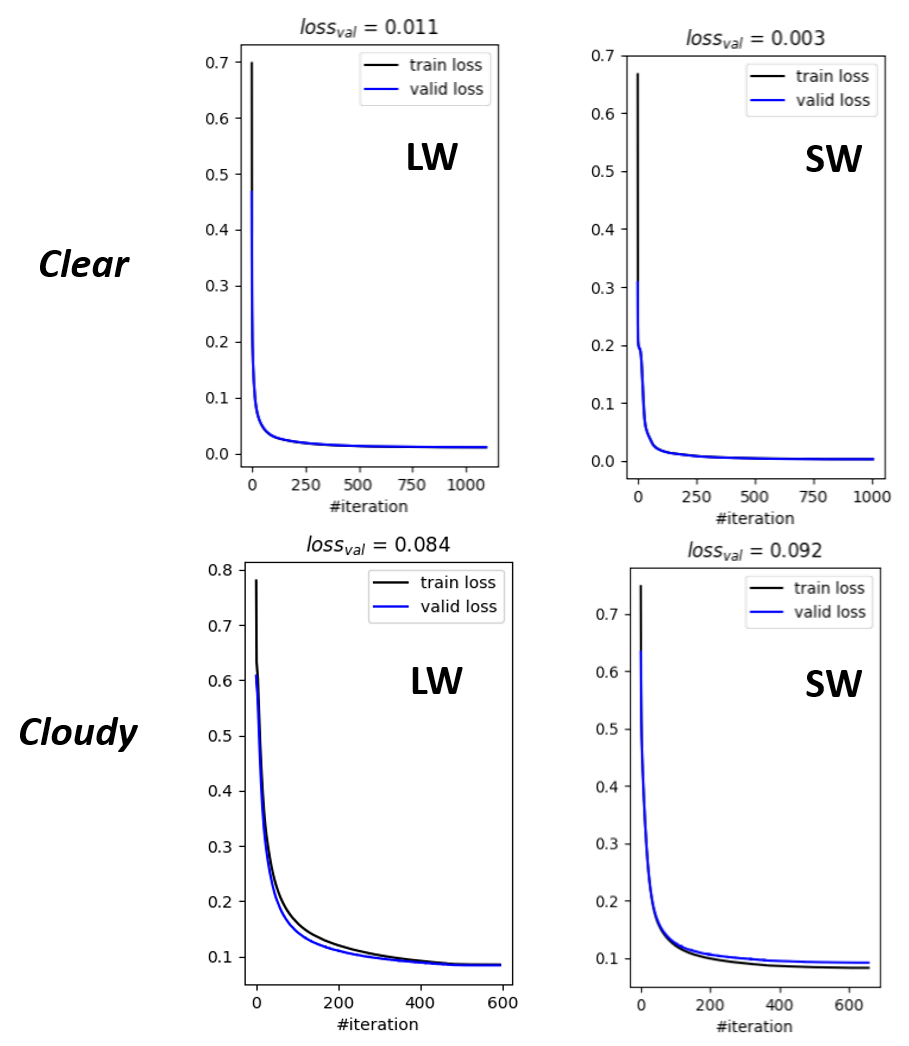}}
	\caption{Convergence of the AI models for clear and cloudy conditions, longwave (LW) and short wave (SW).}
	\label{fig:convergence}
\end{figure}

For the given input-output pairs, the neural network method utilized the stochastic neural network (SNN) approach as described by Krasnopolsky \cite{Krasnopolsky2014} . In the Zhejiang Province scenarios, the mean epochs were \textasciitilde 900 under clear conditions and \textasciitilde 700 under cloud conditions, resulting in approximately \textasciitilde 3,600 and \textasciitilde 2,800 epochs utilized, respectively (see Figure \ref{fig:convergence}).

\subsection{Training Pipeline}
\label{subsection:pipeline}
\begin{figure}
	\centerline{\includegraphics[width=3.5in]{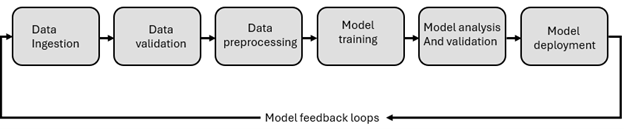}}
	\caption{Understanding the Machine Learning (ML) workflow model. Learning curves for real cases. The results for the sequential neural network (SNN) were derived using the optimal settings identified through subsequent analyses. The optimal epoch and the normalized root mean square error (RMSE) for all outputs are provided in parentheses.}
	\label{fig:pipeline}
\end{figure}

The training set pipeline is a complete set of operations designed to curate, handle, and manage data for machine learning model training, as illustrated in Figure (\ref{fig:pipeline}). Each stage of this pipeline handles a distinct challenge, ensuring that the data used is of high quality, relevance, and usefulness. The successful execution of this pipeline is critical to the creation of strong and trustworthy machine learning models. As part of the first phase of the project, training data was produced for the Zhejiang Province using the original WRF model. 

Training datasets for the Zhejiang Province were then generated by random sampling from the comprehensive dataset that was created in the first phase. The number of samples varied, ranging from approximately 250,000 for clear sky models to around 400,000 for cloudy models. This sampling was carried out based on 30-minute interval outputs simulated for the months of September, between 2015 to 2021, using the traditional WRF model.
The training data was then filtered into eight distinct categories based on combinations of clear/cloudy conditions, longwave/shortwave radiation, and ocean/land surfaces. This filtering allowed each model to focus on specific atmospheric conditions, thus potentially improving accuracy \cite{Song2021}. After filtering, the data was uniformly sampled in space with deterministic 30-minute time intervals. The data was split into 90\% for training and 10\% for validation. Additional data from simulations of September 2022 was used to test the model. 

\subsection{Data Distribution}
\begin{figure}
	\centerline{\includegraphics[width=3.5in]{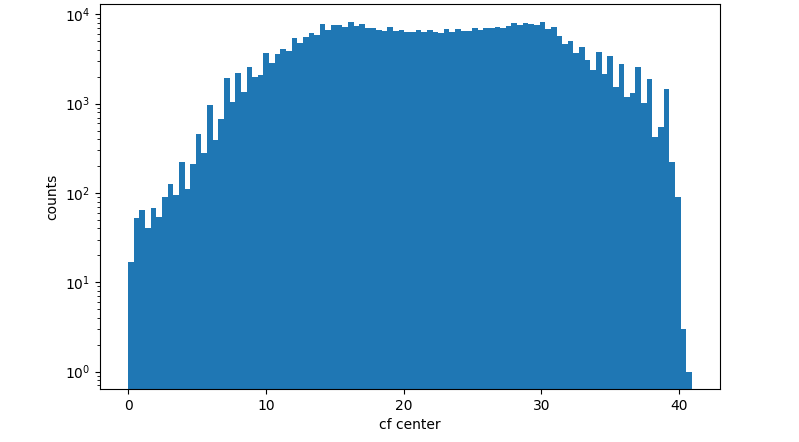}}
    \caption{Histogram showing the distribution of the cloud fraction center of mass for 412,704 training samples of a cloudy model. Noticeable declines in sample counts at low and high center of mass values underscore the necessity of dividing the data into folders as described to ensure that models are trained with a diverse range of cloud scenarios, enhancing their predictive accuracy across different atmospheric conditions}
    
    \label{fig:cloud_fraction_distribution}
\end{figure}

We propose a two-step methodology to represent the structure of layered clouds in the atmosphere, aimed at improving sampling accuracy and ensuring balanced distribution, which is critical for accurate radiation modeling. The first step involves classifying cloud cover over all pixels (longitude/latitude) within the domain into 10 categories, based on the temporal average of cloud fraction and elevation. In the second step, these categories are further subdivided into 10 additional groups, using the center of mass, calculated as the weighted average of cloud fraction across multiple vertical (Z) levels. As shown in Figure (\ref{fig:cloud_fraction_distribution}), the pronounced difference between the central values and the tails of the cloud fraction distribution necessitates this approach to prevent the under-representation of edge cases, particularly at the extremes of low and high cloud fraction. 

\subsection{Implementation within WRF}
\label{subsection:inference}

The proposed single-layer neural network (NN) emulator (see section \ref{subsection:NNModel}) was integrated into the WRF model function $\textit{module\_ra\_rrtmg\_swk.F}$. The modified WRF model is hereafter referred to as WRF-NN.
For the sake of implementation, the NN RRTMG sub-models were organized into three groups (lwr) based on land use (l), weather (w), and radiation modes (r). The land use is categorized as either Land (L) or Ocean (O). The weather types are defined by whether the conditions are Clear (1) or Cloudy (2). Finally, The radiation is classified as either Longwave (L) or Shortwave (S) \footnote{Note that the Long- and Short-wave radiation modes are already separated in the original design of WRF, since short-wave radiation is only used during the daytime when the sun is up.}.
For instance, the "L1S" model represents the shortwave model for clear land.

This NN emulator produces approximated outputs for given inputs to the WRF RRTMG function, circumventing the complex processes of the original parametrization. The model inference followed the formulation given by:
\begin{equation}\label{eq:equation1}    
\begin{split}
Y_q &= B2_q + \sum_{j=1}^{k} W2_{qj} \cdot \tanh\left(
    B1_j + \sum_{i=1}^{n} W1_{ji} \cdot X_i
\right), \\
&\quad q = 1, 2, \ldots, m
\end{split}
\end{equation}

In this context, $n$ and $m$ denote the number of inputs and outputs, respectively; $X_i$ and $Y_q$ represent the input and output vectors; $W1$ and $W2$ are the weight matrices from the input to hidden layers [$n$,$k$]  and from the hidden to output layers [$k$,$m$], respectively; and $B1$ and $B2$ are the bias vectors for the input to hidden layers and hidden to output layers, respectively. The vertical dependencies between input and output variables are adjusted using the weight and bias coefficients. The accuracy of the NN emulator may be further enhanced by increasing the number of hidden neurons $k$, on account of added computational complexity.

\section{Results and Discussions}
\label{section:resultsanddiscussions}

\subsection{Typhoon Muifa}
\label{section:studyarea}
Zhejiang Province, often referred to as the "Land of Fish and Rice," is a captivating blend of natural beauty and cultural richness in southeastern China. At the heart of Zhejiang's meteorological profile is its subtropical climate, which gifts the province with mild winters and sultry summers. However, amidst its serene beauty lies a formidable force of nature: typhoons. These powerful tropical storms, originating from the warm waters of the Pacific Ocean, frequently make landfall along Zhejiang's coastline, bringing heavy rainfall, strong winds, and occasional flooding. Despite the challenges they pose, typhoons are an integral part of Zhejiang's ecosystem, replenishing its water reservoirs, shaping its coastal geography, and influencing its agricultural cycles. 
Autumn, spanning September through November, is generally characterized by dry, warm, and clear sky weather, with average temperatures ranging from 16 to 21°C. July claims the title of the hottest month, with average temperatures hovering between 26 and 29°C.
Throughout the summer and early autumn, typically from June to October, Zhejiang Province is prone to typhoon activity, with August and September experiencing the highest frequency. These tropical cyclones bring strong winds, heavy rainfall, and the risk of flooding to Zhejiang and other coastal regions of China. 
In the current investigation, the simulation conducted in September 2022 coincided with the onset of Typhoon Muifa. On Wednesday, September 14, 2022, strong winds and heavy rainfall battered China's densely populated east coast. By Friday, September 16, 2022, the storm had intensified, with winds accelerating to speeds of up to 82 kilometers (50 miles) per hour.
Following the devastating impact of the Muifa typhoon in 2022, the necessity for fast and precise weather forecasting in Zhejiang Province became increasingly apparent. As communities grappled with the aftermath of the storm, the importance of accurate predictions to anticipate and mitigate future natural disasters became paramount.

\begin{figure}
	\centerline{\includegraphics[width=2.8in]{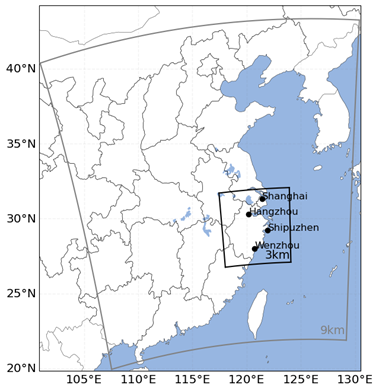}}
	\caption{Domain configuration for Zhejiang Province: - Reference Coordinates: Latitude 29, Longitude 123.5 - Horizontal Grid Projection: Lambert Conformal Conic - Grid dimensions: 255x225 and 370x370 points in the x and y directions - Grid spacing: 9 km and 3 km in both x and y directions.}
	\label{fig:domain}
\end{figure}
The Weather Research and Forecasting (WRF) model version 4.5.1 \cite{Skamarock2008} with two domains (see Figure \ref{fig:domain}) is utilized to perform a 7-day simulation. The model has 45 vertical layers in the terrain-following sigma ($\sigma$), distributed from the surface up to 5 hPa. The horizontal grid spacing of the two domains is 9 km and 3 km, respectively (see Figure \ref{fig:domain}). The time step is 30 s, and the output interval is 1 h. The model employs the WRF Single-Moment (WSM) 6-class graupel micro-physical scheme \cite{Hong2006}, Yonsei University boundary layer scheme ( \cite{Noh2003}; \cite{Hong2006}; \cite{Hong2008} ), Noah land surface model ( \cite{Chen2001}; \cite{Ek2003} ), Monin Obukhov surface layer scheme \cite{Monin1954}, rapid radiative transfer model longwave radiation scheme \cite{Mlawer1997}, and Dudhia shortwave radiation scheme \cite{Dudhia1989}. The NCEP Global Forecast System (GFS) Analyses and Forecasts data are used as the initial and boundary values of the WRF in this study. It has a horizontal resolution of 0.25° × 0.25° and updates every six hours from 00:00 to 18:00UTC.

\subsection{Vortex Tracking}
\label{subsection:vortex_tracking}
\begin{figure}
	\centerline{\includegraphics[width=3.5in]{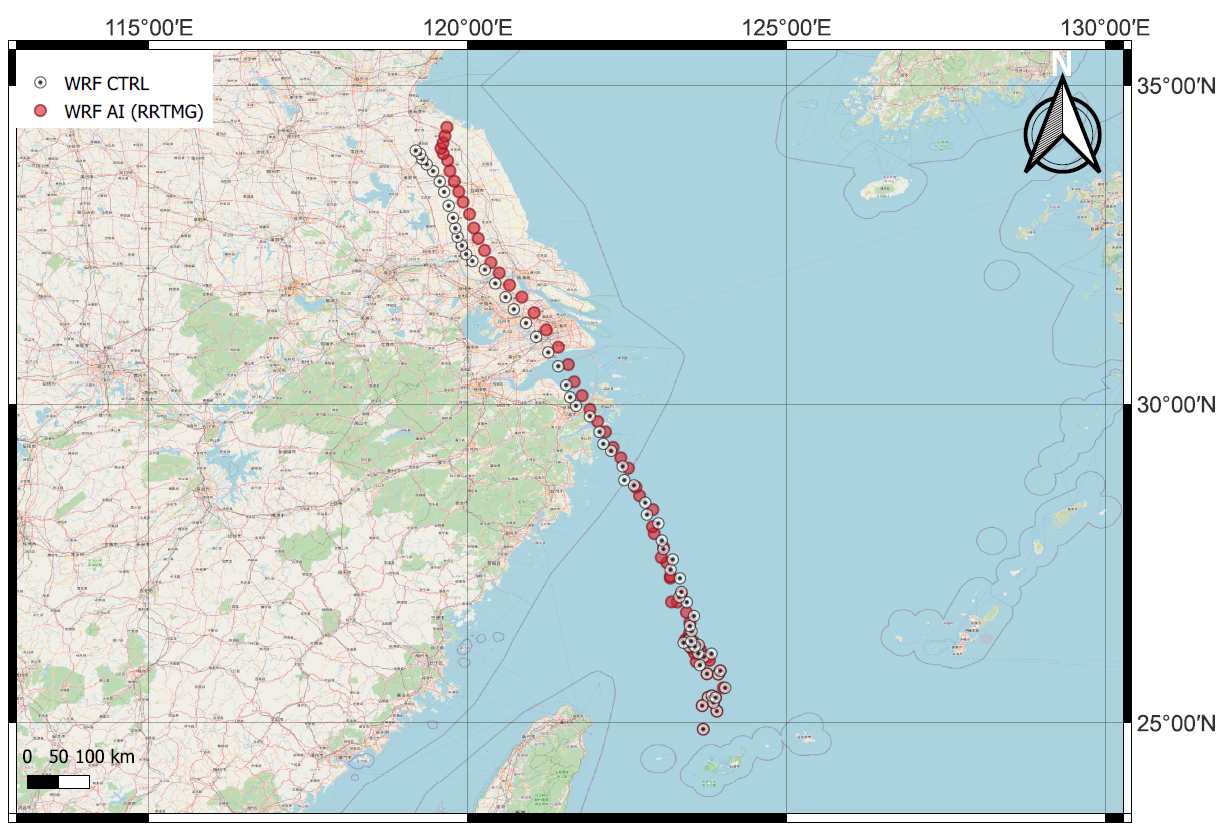}}
	\caption{Simulation duration of Typhoon Muifa by the WRF-RRTMG and WRF-NN models. The WRF-RRTMG track is represented by black dots, and the WRF-NN track is depicted by red dots.}
	\label{fig:nnvortextracking}
\end{figure}

The results of the WRF-NN model are compared with the WRF-RRTMG model track of the Muifa Typhoon to evaluate its capability on Tropical Cyclone (TC) prediction. Center of Typhoon was accurately computed by looking for the location of minimum pressure and minimum wind speed.  Figure \ref{fig:nnvortextracking} shows the 7 days track of typhoon Muifa simulated by WRF-RRTMG and WRF-NN. Both simulations (WRF-RRTMG and WRF-NN) show typhoon location in agreement. Towards the end of the simulation, the results diverge slightly, mostly due to proximity of domain boundaries

\subsection{Comparative Analysis of Controlled and AI Simulations}
\label{subsection:comparative_analysis}
The Pearson correlation coefficient analysis \cite{Pearson1895} involves carefully comparing two simulations: one controlled (CTRL) and the other using the surrogate RRTMG (AI) model. The primary aim is to rigorously evaluate the disparity between the outcomes produced by each simulation method. By meticulously analyzing these metrics, we collect profound insights into the performance and efficacy of the AI simulation in contrast to the conventional CTRL simulation. This experiment, designed to uphold principles of fairness and transparency, aims to furnish robust evidence concerning the RRTMG surrogate model's aptitude in simulating the given scenario.

\subsection{Temporal Validation and Error Analysis}

Accurate simulation of surface air temperature (T2) and skin temperature (Tskin) is crucial for understanding atmospheric processes, especially during extreme weather events such as typhoons. To assess the performance of the Inference NN surrogate model, we conducted a temporal validation by comparing its predictions against the CTRL WRF simulation for the Mufia Typhoon event. The evaluation focuses on the root mean square error (RMSE) of both T2 and Tskin across the spatial domain.

\begin{figure}
	\centerline{\includegraphics[width=3.5in]{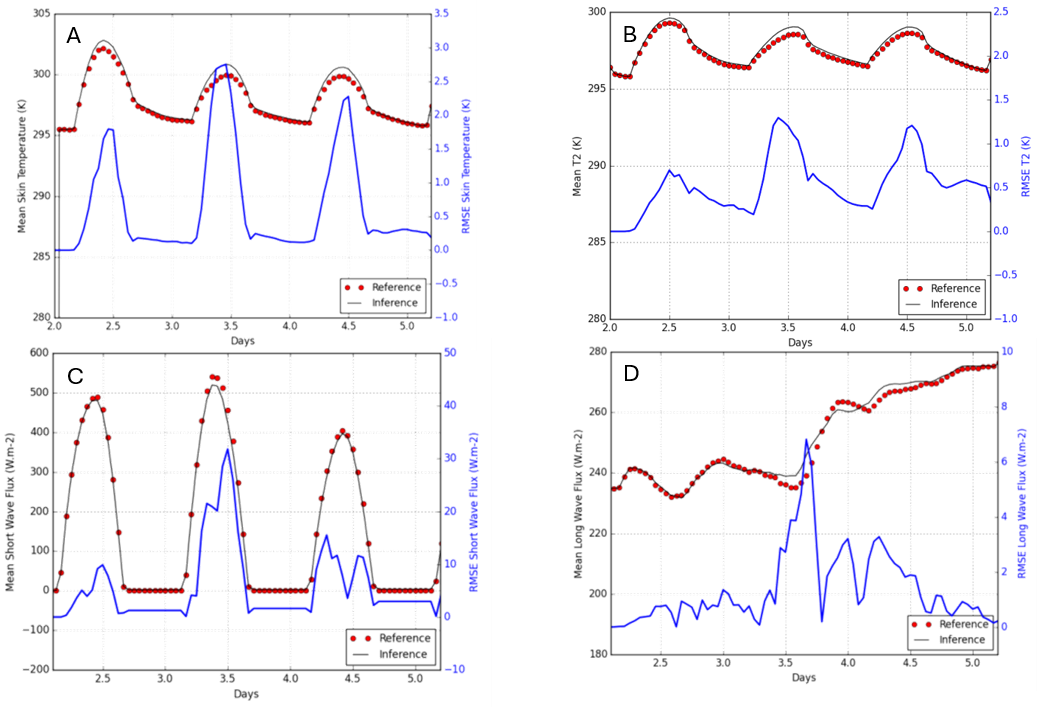}}
	\caption{Temporal validation of mean skin temperature in Panel (A), mean surface air temperature in Panel (B), shortwave radiation in Panel (C), and longwave radiation in Panel (D) across the domain as a function of time.}
	\label{fig:temporal_validation_and_error_analysis}
\end{figure}

The validation results, as illustrated in the Figure \ref{fig:temporal_validation_and_error_analysis}, show that the absolute error remains below 2K for Tskin and under 1.3K for T2. A distinct 12-hour oscillation in error magnitude is observed, with a peak around mid-day, highlighting the influence of shortwave radiation—primarily solar radiation reaching the Earth's surface—on temperature discrepancies. This mid-day peak suggests that the simulation's representation of solar energy flux may contribute to these errors.

Additionally, a notable increase in longwave radiation is observed after three and a half days, followed by an exponential decrease. This sudden rise in longwave radiation, which consists mainly of infrared radiation emitted by the Earth's surface and the atmosphere, plays a significant role in influencing T2 and Tskin. The increase indicates a transition in atmospheric conditions, which may enhance heat retention in the lower atmosphere. The subsequent exponential decrease may reflect the dissipative effects of the typhoon's dynamics as it progresses, impacting temperature profiles and contributing to fluctuations in the error metrics.

Despite these periodic variations and the influence of both shortwave and longwave radiation, the surrogate model demonstrates high accuracy, with an overall error of less than 0.6\% for both T2 and Tskin.

\subsection{Spatial Accuracy}
Spatial accuracy is a critical metric in meteorological forecasting, particularly when utilizing neural network (NN) surrogate models in place of traditional numerical models such as the Weather Research and Forecasting (WRF) model. Accurate representation of the spatial distribution of meteorological variables—such as temperature, pressure, wind fields, and precipitation—is essential for both short-term weather predictions and long-term climate modeling.

In meteorology, weather phenomena are inherently spatial and exhibit complex, non-linear patterns over varying scales. These include localized events, such as thunderstorms or tornadoes, as well as larger-scale systems, such as hurricanes or atmospheric fronts. Any degradation in spatial accuracy can lead to misrepresentations of these critical features, reducing the forecast model's overall effectiveness. 

The WRF model is well-known for its ability to provide detailed weather data over both space and time, which makes it highly effective for capturing small-scale weather patterns.
NN surrogates, which aim to reduce the computational demands of running full-scale numerical models, must therefore maintain a high level of spatial precision to be effective. High spatial accuracy ensures that the surrogate model captures localized weather phenomena, a necessity for sectors such as agriculture, disaster management, and energy, where detailed, high-resolution forecasts are critical. Without such accuracy, the surrogate model's utility would be significantly diminished.

\begin{figure}
	\centerline{\includegraphics[width=3.5in]{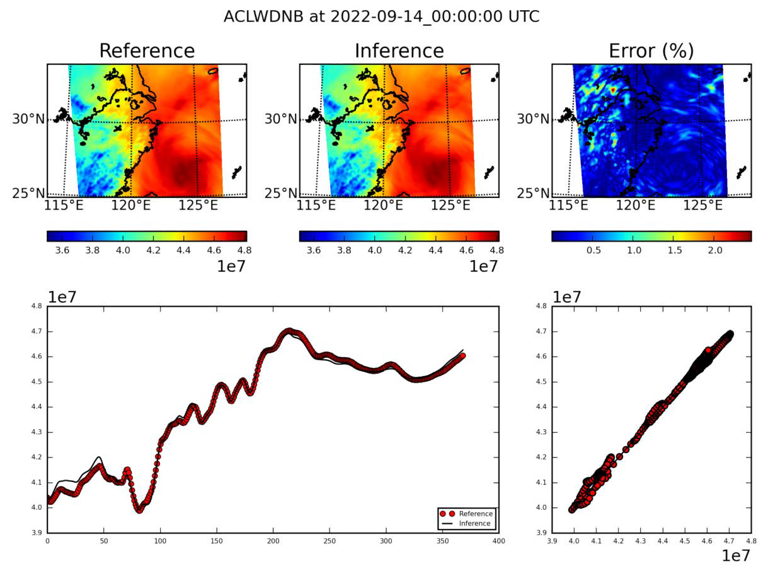}}
	\caption{Spatial accuracy of downward LW radiation on 2022/09/14 00:00:00 UTC. The top row displays (from left to right) results from the traditional WRF model, the corresponding inference from the NN surrogate model, and the spatial error in percentage. The bottom row shows (i) a profile scan across the entire domain, and (ii) a scatter plot showing minor deviation from the ideal 1:1 line (black).}
	\label{fig:spatial_validation_and_error_analysis1}
\end{figure}

\begin{figure}
	\centerline{\includegraphics[width=3.5in]{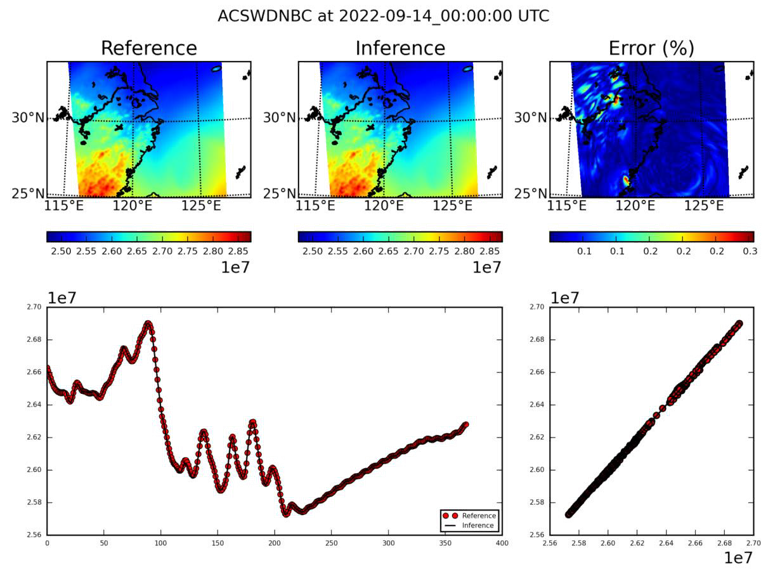}}
	\caption{Spatial accuracy of downward SW radiation on 2022/09/14 00:00:00 UTC. The top row displays (from left to right) results from the traditional WRF model, the corresponding inference from the NN surrogate model, and the spatial error in percentage. The bottom row shows (i) a profile scan across the entire domain, and (ii) a scatter plot showing minor deviation from the ideal 1:1 line (black).}
	\label{fig:spatial_validation_and_error_analysis2}
\end{figure}

\begin{figure}
	\centerline{\includegraphics[width=3.5in]{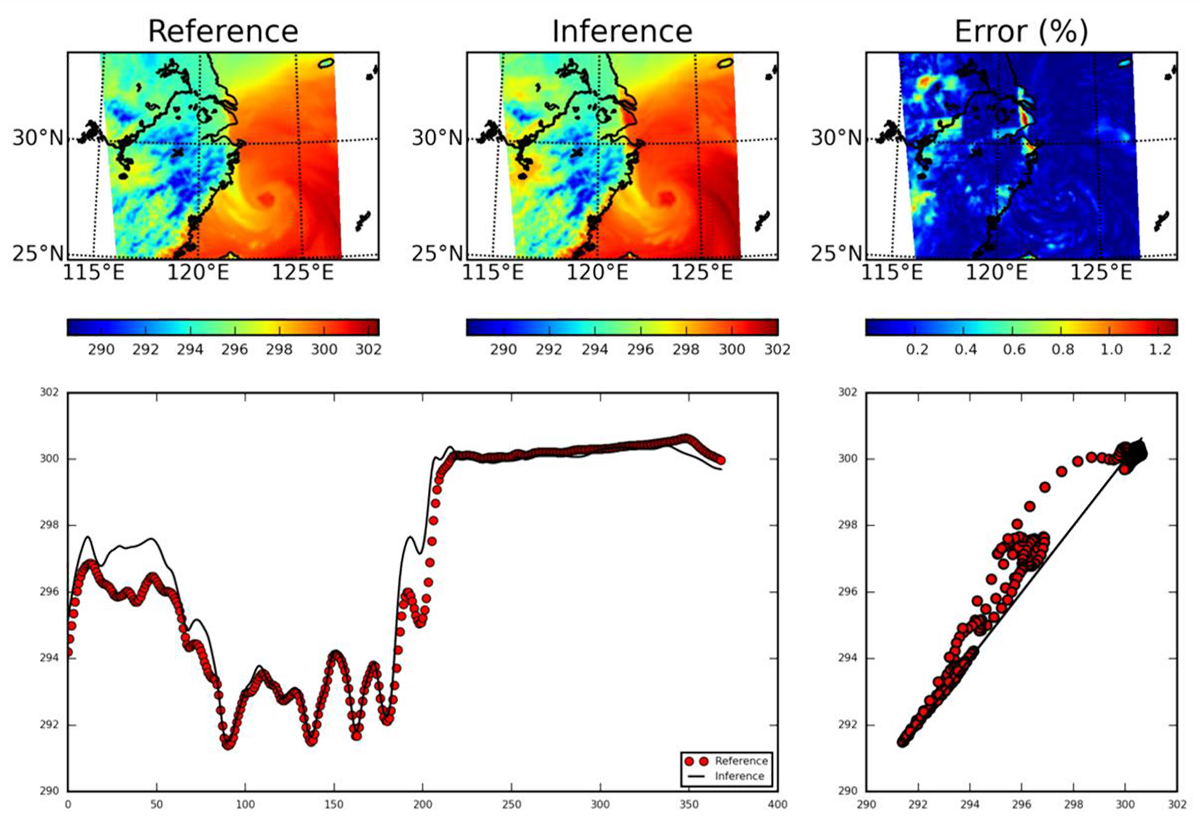}}
	\caption{Spatial accuracy of 2-meter temperature on 2022/09/14 00:00:00 UTC. The top row displays (from left to right) results from the traditional WRF model, the corresponding inference from the NN surrogate model, and the spatial error in percentage. The bottom row shows (i) a profile scan across the entire domain, and (ii) a scatter plot showing minor deviation from the ideal 1:1 line (black).}
	\label{fig:spatial_validation_and_error_analysis3}
\end{figure}

The spatial accuracy of the NN surrogate model is highlighted in Figures \ref{fig:spatial_validation_and_error_analysis1}, \ref{fig:spatial_validation_and_error_analysis2} and \ref{fig:spatial_validation_and_error_analysis3}, which present a comparison of the results field between the NN surrogate and the reference WRF model. The results in Figure \ref{fig:spatial_validation_and_error_analysis3} demonstrate strong spatial agreement, with the surrogate model replicating the temperature distribution with high fidelity. Importantly, the maximum error is localized to a small region where the deviation peaks at 1.2\%. In contrast, the majority of the domain exhibits errors of less than 0.2\%, underscoring the model's exceptional spatial accuracy and reliability in predicting near-surface temperature fields. Similar results can be seen when looking at the short wave and long wave predictions.

These findings emphasize that improving spatial accuracy in NN surrogate models is not merely an enhancement but a necessity to maintain the reliability of weather forecasts. Accurate spatial predictions empower meteorologists and decision-makers to better prepare for extreme weather events, reduce risk, and manage resources effectively.

\subsection{Pearson Coefficient Correlation for the Typhoon Muifa}

To further evaluate the accuracy of the developed approach, the Pearson Coefficient Correlation was computed for multiple variables of interest. The Pearson Coefficient is used to measure the correlation between two sets of results and is defined as the ratio between the covariance of the two sets and the product of their standard deviations. It can be written as:

\begin{equation}
    r=\frac{n\sum x_iy_i - \sum x_i\sum y_i}{\sqrt{n\sum x_i^2 -(\sum x_i)^2}\sqrt{n\sum y_i^2 -(\sum y_i)^2}}
\end{equation}

where $n$ is the sample size, $x_i$ and $y_i$ are the individual points of each dataset. In our case, we are comparing the results of the WRF predictions without AI model, with the results obtained with the RRTMG AI surrogate model. 
\begin{table}[!ht]
    \centering
    \begin{tabular}{||c|c||}
    \hline
       Variable & Pearson Coefficient \\ \hline
       ACSWDNBC  & 0.9998172098691426 \\
       ACLWDNB  &  0.9571510899677951 \\
       ACLWUPTC  & 0.9976412955947259  \\
       ACSWDNBC  & 0.9998172098691426  \\
       MU & 0.9977955195732356  \\
       O\_LWDNBC  &  0.9842323196827641 \\
       O\_SWDNBC  &  0.9999466972958478 \\
       T2  &  0.962314874191279 \\
       \hline
    \end{tabular}
    \caption{Pearson Coefficient Correlation between WRF predictions with and without RRTMG AI surrogate model. Data was extracted from the entire computational domain at the end of day 2 of the typhoon forecast.}
    \label{tab:PC}
\end{table}

Table \ref{tab:PC} shows the results obtained at the end of the second day of the forecast. The data were obtained using all the entries from the fine resolution computational domain. The parameters were chosen for their relevance to the radiation physics, and for their known impact of the radiation prediction accuracy. It can be clearly seen that the accuracy of the prediction with AI is excellent, matching the predicted weather forecast from the standard WRF solver. This further confirms the validity of the overall approach, even in the case of extreme event forecast.

\subsection{Computational Speedup}
This subsection provides a quantitative comparison between the traditional RRTMG model within WRF and the AI-based WRF surrogate model, 
focusing on their iteration times and computational efficiency under different weather conditions. 

The WRF model shows considerable variation in iteration times depending on meteorological conditions, such as clear sky and cloudy weather. 
In contrast, the AI-based WRF surrogate model maintains stable iteration times, resulting in improved efficiency and stability.

\begin{figure}
	\centerline{\includegraphics[width=3.5in]{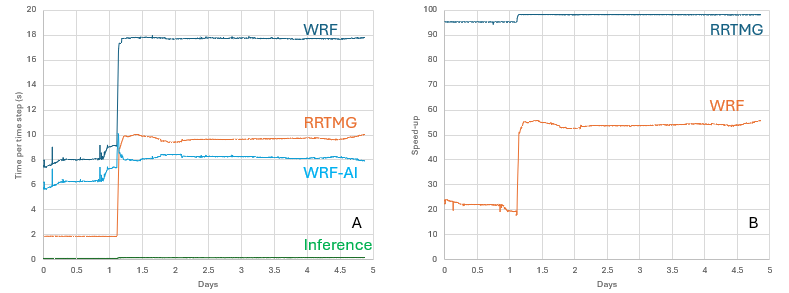}}
	\caption{Panel (A) illustrates the wall time per time step throughout the simulation period, while panel (B) presents the speedup achieved during the same simulation time-frame.}
	\label{fig:speedup_over_day}
\end{figure}

During clear sky periods, the RRTMG model requires approximately 2 seconds per iteration, which increases to around 10 seconds during cloudy 
conditions. The AI-WRF surrogate model, however, not only maintains constant iteration times but also reduces radiation 
computation by nearly 99\%. This leads to an overall reduction in total computation time by 20\% on clear sky days and 55\% on cloudy days 
Figure \ref{fig:speedup_over_day}.

This comparison emphasizes the AI inference model’s ability to drastically reduce computational demand, providing a significant advantage over the traditional RRTMG-WRF model, particularly during extreme meteorological events like typhoons.

\subsection{Temporal Transfer Learning for Enhanced Forecasting of Intra-Annual Weather Variability}
\begin{figure}
	\centerline{\includegraphics[width=3.5in]{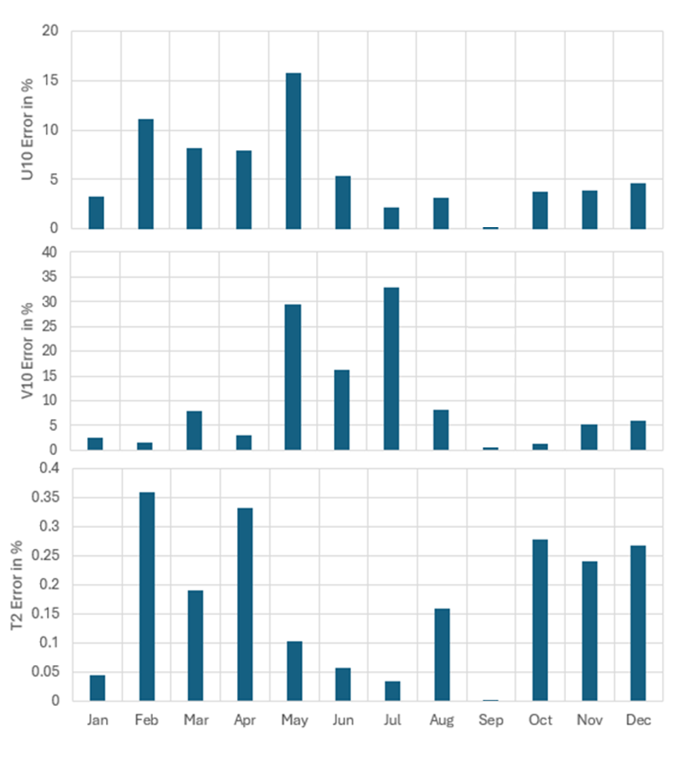}}
	\caption{Mean error per month for the model developed in section III for the region defined in section II. Large error can be seen for months in which the weather conditions are very different from the conditions of the training datasets.}
	\label{fig:Error_per_month}
\end{figure}

As an extension to the month-specific emulation strategy presented above, we further explored how a model trained on one month’s data could be adapted to different months through transfer learning, an AI technique in which a model trained on one dataset is repurposed, with minimal additional training, for a related but distinct dataset. 
We began by training a neural network emulator initialized from a Kaiming uniform distribution on 300,000 September samples. We used a learning rate scheduler (ReduceLROnPlateau from the PyTorch library). that decreases the learning rate once the validation loss fails to improve for 5 consecutive epochs, with a minimum learning rate threshold of 1e-7, which led the training to stop at ~1,500 epochs. This baseline training resulted in an NRMSE of 0.03278 and a loss of 0.006. When we applied this September trained model directly to other months, performance degraded considerably, reflecting the climatological discrepancies between different times of the year, as shown in Figure \ref{fig:Error_per_month}.
To address this, we fine-tuned the model’s weights for July by initializing with the September trained parameters and then training on 100,000 new July samples. Again, we employed early stopping based on validation loss, and this time the network converged after ~400 epochs - significantly fewer than the original 1,500. Notably, the transfer learning step reduced the NRMSE to 0.01037 and the loss to 0.006, demonstrating that previously learned relationships, such as how temperature or pressure profiles affect radiative transfer,accelerate and improve adaptation to new conditions. The results are summarized in Table \ref{table:retraining}.
Our approach not only reduced the final error metrics but also required fewer epochs to achieve them and a smaller data set, showcasing the computational efficiency of transfer learning within our weather forecasting setup. Moreover,


This successful month-to-month knowledge transfer highlights the potential to extend beyond individual monthly models.


If further refined, our method could facilitate the development of a single year-round emulator capable of adapting to a wide range of atmospheric scenarios, significantly reducing the workload associated with maintaining separate emulators for each month.

\begin{table}
\begin{center}
\begin{tabular}{|| c | c | c ||}
\hline
 & Fresh training & Transfer learning \\
\hline
 Number of Training samples & 300000 & 100000 \\ 
 \hline
 Number of Training epochs & 1500 & 400 \\  
 \hline
 Final error (NRMSE) & 0.03278 & 0.01037 \\
 \hline
\end{tabular}
\caption{Cost saving of the transfer learning when compared to fresh model training}
\label{table:retraining}
\end{center}
\end{table}

\section{Conclusions}
\label{section:conclusions}
To meet the urgent need for fast and accurate weather forecasting, the development and implementation of advanced forecasting tools, such as the proposed WRF-RRTMG AI surrogate model, are of top priority. 
This cutting-edge model was designed to accelerate weather forecasts with the WRF model, demonstrating up to 2x faster simulation of the challenging Typhoon Muifa event that stormed China's Zhejiang Province in 2022, with minimal impact on accuracy.

Despite being trained specifically for forecasting this extreme event, the proposed model was also shown to support accurate forecast of atmospheric conditions over the course of the entire year thanks to the temporal transfer learning approach. 

The WRF-RRTMG surrogate model was successfully implemented in the WRF framework, to  provide invaluable insights into the behavior and trajectory of approaching weather systems, including typhoons.

Finally, this work showcases how harnessing the power of advanced computational algorithms and real-time data assimilation, meteorologists can now deliver timely warnings and alerts to vulnerable communities, allowing them to better prepare and respond to potential threats.


\section{Acknowledgments}
Professor Erick Fredj deeply thanks Dr. Soonyoung Roh from the National Institute of Meteorological Sciences in Korea's Meteorological Administration for her assistance during the initial phase of this work. 

\bibliographystyle{IEEEtran}
\bibliography{IEEEabrv,./arxiv_2024_bib}
\end{document}